\documentstyle[prb,aps,multicol,epsf]{revtex}
\begin{document}
\title{Negative Pressure of Anisotropic Compressible Hall States : 
Implication to Metrology} 
\author{Kenzo Ishikawa and Nobuki Maeda}
\address{
Department of Physics, Hokkaido University, 
Sapporo 060-0810, Japan}
\date{\today}
\maketitle
\begin{abstract}
Electric resistances, pressure, and compressibility of anisotropic 
compressible states at higher Landau levels are analyzed. 
The Hall conductance varies continuously with filling factor 
and longitudinal resistances have huge anisotropy. 
These values agree with the recent experimental observations 
of anisotropic compressible states at the half-filled higher Landau levels. 
The compressibility and pressure become negative. 
These results imply formation of strips of the compressible gas which 
results in an extraordinary stability of the integer quantum 
Hall effect, that is, the Hall resistance is quantized exactly even when the 
longitudinal resistance does not vanish.  
\end{abstract}
\draft
\pacs{PACS numbers: 73.40.Hm, 73.20.Dx}

\begin{multicols}{2}

Recently highly correlated anisotropic states have been 
observed around half-filled higher Landau levels of high mobility GaAs/AlGaAs 
hetero-structures.\cite{a,b,d,e} 
Longitudinal resistance along one direction tends to vanish at low temperature 
but that of another direction has a large value of order K$\Omega$. 
The Hall resistance is approximately proportional to the filling factor. 
Since one longitudinal resistance is finite, the state is compressible. 
It has been noted also that the current-induced breakdown and collapse of 
the quantum Hall effect (QHE) at $\nu=4$ occurs through several steps 
which implies that compressible gas in quantum Hall system has unusual 
properties.\cite{f,g} 

Anisotropic stripe states were predicted to be favored in higher Landau 
levels.\cite{h,i} 
Using the von Neumann lattice basis, the present authors found an 
anisotropic mean field state in the lowest Landau level to have a negative 
pressure and negative compressibility.\cite{j,eisen} 
Their energies at higher Landau levels were calculated recently by 
one of the present authors\cite{k} and others\cite{l} in the 
Hartree-Fock approximation. 
They have lower energies than symmetric states, but their physical properties 
have not been studied well so that it is not clear 
if these states agree with the states found by experiments.

In the present work, we study the physical properties of anisotropic mean 
field states around $\nu=n+1/2$ and $\nu=n$, where $n$ is an integer. 
We point out that this mean field state at $\nu=n+1/2$ has properties 
described above and can explain the experimental observations of anisotropic 
states. 
These states have a negative pressure and negative compressibility and 
periodic density modulation in one direction. 
We show that as a consequence of negative pressure, 
a compressible gas strip is formed in the bulk around $\nu=n$ and 
a current flows in the strip by a new tunneling mechanism, 
activation from undercurrent. 
The current is induced in isolated strip with a temperature dependent 
magnitude of activation type and it causes a small longitudinal resistance 
in the system of quantized Hall resistance. 
This solves a longstanding puzzle of the integer QHE, namely Hall 
resistance is quantized exactly even if the system has a small finite 
longitudinal resistance. 
Collapse phenomena\cite{f} are shown to be understandable also. 

The von Neumann lattice basis\cite{m,n} is one of the bases for 
degenerate Landau levels of the two-dimensional continuum space where 
discrete coherent states of guiding center variables $(X,Y)$ are used and 
is quite useful in studying QHE because translational 
invariance in two dimensions is preserved. 
Spatial properties of extended states and interaction effects were studied 
in systematic ways. 
We are able to express exact relations such as current conservation, 
equal time commutation relations, and Ward-Takahashi identity
\cite{o} in equivalent 
manners as those of local field theory and to connect the Hall conductance 
with momentum space topological invariant. 
Exact quantization of the Hall conductance in quantum Hall regime (QHR) 
in the systems of disorders, interactions, and finite injected current 
has been proved in this basis. 
We use this formalism in the present paper as well.

Electrons in the Landau levels are expressed with the creation and 
annihilation operators 
$a_l^\dagger({\bf p})$ and $a_l({\bf p})$ of having Landau level index, 
$l=$0, 1, 2$\dots$, and momentum, $\bf p$. 
The momentum conjugate to von Neumann lattice coordinates is defined in the 
magnetic Brillouin zone (MBZ), 
$
\vert p_i\vert \leq{\pi/a},\ a=\sqrt{2\pi\hbar/eB}. 
$
The many body Hamiltonian $H$ is written in the momentum representation as 
$H=H_0+H_1$, where
\begin{eqnarray}
H_0&=&\sum_{l=0}^{\infty}\int_{MBZ} {d{\bf p}\over (2\pi/a)^2} 
E_l a_l^\dagger({\bf p})a_l({\bf p}),\label{ham}\\
H_1&=&\int_{{\bf k}\neq0} 
d{\bf k}\rho({\bf k}){V({\bf k})\over2}\rho(-{\bf k}).\nonumber
\end{eqnarray}
Here $E_l$ is the Landau level energy $(\hbar eB/m)(l+1/2)$ and 
$V({\bf k})=2\pi q^2/k$ for the Coulomb interaction, and the charge 
neutrality is assumed. 
In Eq.~(\ref{ham}), $H_0$ is diagonal but the charge density 
$\rho({\bf k})$ is non-diagonal with respect to $l$. 
We call this basis the energy basis. 
A different basis called the current basis in which charge density becomes 
diagonal will be used later in computing current correlation functions 
and electric resistances. 

It is worthwhile to clarify the peculiar symmetry of the system described by 
Eq.~(\ref{ham}). 
The Hamiltonian is invariant under translation in momentum space, 
${\bf p}\rightarrow {\bf p}+{\bf K}$, where $\bf K$ is a constant vector, 
which is called the $K$-symmetry. 
This symmetry emerges because the kinetic energy is quenched due to the 
magnetic field. 
A state which has  momentum dependent single particle energy violates 
the $K$-symmetry.\cite{ks} 
In the present paper, we study a mean field solution which violates 
$K_y$-symmetry but preserves $K_x$-symmetry. 
The one-particle energy has $p_y$ dependence in this state.  

The compressible gas state is characterized by the following form of 
expectation values in the coordinate space, 
\begin{equation}
U^{(l)}({\bf X}-{\bf X}')\delta_{ll'}=\langle a^\dagger_{l'}({\bf X}')a_l
({\bf X})\rangle,
\end{equation}
where the expectation values are calculated self-consistently 
in the mean field approximation\cite{j,k} using $H_1$ and the mean 
field  $U^{(l)}$. 
In Figs.~1 and 2, the energy per particle, pressure, 
and compressibility are presented with respect to the filling factor 
$\nu=n+\nu'$. 
As seen in these figures, they become negative. 
The density is uniform in y-direction but is periodic in 
x-direction.\cite{k} 
The present anisotropic state could be identified with the stripe structure 
discussed in Refs.~\cite{h} and \cite{i}. 
We have checked that bubble states discussed in Ref.~\cite{h} 
also have negative pressure and compressibility. 
These properties may be common in the compressible states of the quantum Hall 
system.\cite{efros} 

The compressible states thus obtained have negative pressure and are 
different from ordinary gas. 
Naively it would be expected that these gas states were unstable. 
However thanks to the background charge of dopants, a stable state with 
a negative pressure can exist. 
Since the pressure is negative, charge carriers compress itself while 
dopants do no move. 
Then charge neutrality is broken partly and compressible gas states are 
stabilized by Coulomb energy. 
Total energy becomes minimum with a suitable shape which depends on the 
density of compressible gas. 

The bulk compressible gas states are realized around $\nu=n+1/2$ 
(called region I), where the Coulomb energy is dominant over negative 
pressure and a narrow depletion region is formed at the boundary. 
Its width is determined by the balance between the pressure 
and the Coulomb force. 

The low density compressible gas states are realized around $\nu =n$ 
(called region II). 
In this case pressure effect is enhanced relatively compared to Coulomb 
energy and a strip of compressible gas states is formed as 
shown in Fig.~3 (a). 
Real system has disorders by which most electronic states are localized. 
Let us classify three different regions depending on the relative ratio 
between localization length $\xi$ at the Fermi energy 
and the width at potential probe area $L_p$ and the width at Hall probe 
area $L_h$. 
We assume $L_p < L_h$. 
In the region II-(i), $\xi < L_p$ is satisfied and localized states fill 
whole system. 
In the region II-(ii), $L_p < \xi < L_h$ is satisfied and 
the Hall probe area is filled with localized states but potential probe 
area is filled partly with compressible gas strip. 
Finally in the region II-(iii), $L_h < \xi$ is satisfied and 
whole area are filled partly with compressible gas states. 
In each case if localization length is longer than the width of the system, 
then these localized states are regarded as extended states 
which behave like compressible gas states with a negative pressure. 
In the regions II-(ii) and II-(iii), the strip contributes to electric 
conductance if current flows through the strip. 
However the strip is unconnected with source drain area. 
How does the current flow through the strip? 
This problem has not been studied before. 
In these regions, extended states below Fermi energy 
carry non-dissipative current, which we call the undercurrent. 
See Fig.~3 (b). 
We show later that the undercurrent actually induces the dragged current. 

First we calculate the electric conductance of the bulk compressible states 
in the region I. 
It is convenient to use current basis for computing current correlation
functions. 
Field operators $a_l$ and propagator $S_{ll'}$ are transformed 
from the energy basis to the current basis as,
\begin{eqnarray}
\tilde a_l({\bf p})&=&\sum_{l'}U_{ll'}({\bf p})a_{l'}({\bf p}),\\
\tilde S_{ll'}(p)&=&\sum_{l_1 l_2}
U_{ll_1}({\bf p})S_{l_1l_2}(p)U^\dagger_{l_2l'}({\bf p}),
\nonumber
\end{eqnarray}
where $U({\bf p})=e^{-i p_x\xi}e^{-ip_y\eta}$ and 
$(\xi,\eta)$ are relative coordinates defined by $(x-X,y-Y)$. 
In the current basis, the equal time commutation relation between the 
charge density and the field operators are given by,
\begin{equation}
[\rho({\bf k}),\tilde a_l({\bf p})]=-\tilde a_l({\bf p})
\delta^{(2)}({\bf p}-{\bf k}). 
\end{equation}
Hence vertex part is given by a derivative of inverse of the propagator,
$\tilde \Gamma_\mu(p,p)=\partial_\mu\tilde S^{-1}(p)$, 
known as Ward-Takahashi identity.\cite{o} 
The Hall conductance is the slope of the current-current correlation 
function at the origin and is given by the topologically invariant 
expression of the propagator in the current basis as\cite{m,n}
\begin{equation}
\sigma_{xy}={e^2\over h}{1\over24\pi^2}\int{\rm tr}({\tilde S}(p)
d{\tilde S}^{-1}(p))^3. 
\label{toppo}
\end{equation}
This shows that $\sigma_{xy}$ is quantized exactly in QHR 
where the Fermi energy is located in the localized state region. 
Now Fermi energy is in the compressible state band region and $\sigma_{xy}$ 
is not quantized. 
For the anisotropic states, the inverse propagator is given by 
$S^{-1}(p)_{ll'}=\{p_0-(E_l+\epsilon_l(p_y))\}\delta_{ll'}$ where 
$\epsilon_l(p_y)$ is the one-particle energy.\cite{j,k} 
$S(p)$ has no topological singurality and its winding number vanishes. 
Hence the topological property of the propagator in the current basis, 
$\tilde S(p)$, is determined solely by the unitary operator 
$U({\bf p})$, 
and the Hall conductance is written as,
\begin{eqnarray} 
\sigma_{xy}&=&{e^2\over h}{1\over4\pi^2}\int dp\epsilon^{ij}
{\rm tr}\left[S(p)U^\dagger({\bf p})\partial_i U({\bf p})
U^\dagger({\bf p})\partial_j U({\bf p})\right]\nonumber\\
&=&{e^2\over h}(n+\nu').
\label{topo}
\end{eqnarray}
To obtain the final result in the above equation, 
we assumed that the Landau levels are filled completely up to $n$ th level 
and $(n+1)$ th level is filled partially with filling factor $\nu'$. 
The Hall conductance is proportional to the total filling factor.  

The longitudinal conductance in x-direction, $\sigma_{xx}$, 
vanishes since there is no empty state in this direction. 
If a momentum is added in x-direction, one particle should be lifted to 
a higher Landau level. 
There needs a finite energy and $\sigma_{xx}$ vanishes. 
The longitudinal conductance in  y-direction, $\sigma_{yy}$, 
does not vanish. 
One particle energy has a dependence on only $p_y$, 
and the system is regarded as one dimensional. 
One dimensional conductance is given by Buttiker-Landauer formula
\cite{p}. 
We have thus, 
\begin{equation}
\sigma_{yy}={e^2/h},\ \sigma_{xx}=0.
\label{sig}
\end{equation}
The Hall conductance, Eq.~(\ref{topo}), and the longitudinal conductances, 
Eq.~(\ref{sig}), agree with the experimental observations of anisotropic 
states around $\nu=n+1/2\ (n\geq 1)$. 

Next we study the low density region, region II. 
In the first region, II-(i), whole area is filled with localized states, 
hence from the formula Eq.~(\ref{toppo}), the Hall conductance is 
quantized exactly. 
The longitudinal resistances vanish. 
We have 
\begin{equation}
\sigma_{xy}=(e^2/h) n,\ \sigma_{xx}=\sigma_{yy}=0. 
\end{equation}
This corresponds to standard QHR. 

In the regions II-(ii) and II-(iii), a compressible strip bridges one 
edge to the other edge. 
Tunneling combined with an interaction causes the dragged current 
in the strip. 
The conductance due to the tunneling mechanism can be calculated by 
a current-current correlation function shown in Fig.~4. 
The two-loop diagram is the lowest order contribution. 
The dragged current flows in the compressible strip at potential 
probe area in the region II-(ii), and at Hall probe area in the 
region II-(iii). 

In the region II-(ii), the Hall probe area is filled with only 
localized states. 
Hence the Hall conductance is quantized exactly. 
The potential probe area has a finite longitudinal resistance due to 
an electric current in the strip. 
The electric current which flows in the strip makes the strip area to 
have a finite temperature. 
We have thus the exactly quantized Hall resistance and a small longitudinal 
resistance in this region as 
\begin{equation} 
R_{xy}^{-1}=(e^2/h)n,\ R_{xx}^{-1}={(e^2/h)}\varepsilon.
\end{equation}
The small parameter $\varepsilon$ is proportional to the activation form 
$\exp[-\beta(\Delta+mv^2/2)]$, where $\Delta$ is the energy gap 
between the Fermi energy and the lower Landau level, $\beta$ is the 
inverse temperature at strip area, and $v$ is the average velocity of the 
undercurrent states. 
The additional term $mv^2/2$ to $\Delta$ comes from the Galilean boost. 
This region has not been taken into account in the metrology of 
QHE.\cite{q} 

In the region II-(iii), whole area is filled with compressible states. 
Hence from Eq.~(\ref{topo}), the Hall conductance is given by unquantized 
value and the longitudinal resistance becomes finite. 
In this case we have 
\begin{equation} 
R_{xy}^{-1}={(e^2/h)}(n+\varepsilon'),
\ R_{xx}^{-1}={(e^2/h)}\varepsilon.
\end{equation}
That is, QHE is collapsed.\cite{f} 
$\varepsilon'$ has the same temperature dependence as $\varepsilon$. 

The localization length and mobility edge depend on injected current in the 
real system. 
In a small current system, localization lengths are small in a 
mobility gap and corresponds to QHR (region II-(i)). 
In an intermediate current system, they become larger and the strip is 
formed in potential probe area first (region II-(ii)) 
and in Hall probe area second (region II-(iii)). 
QHE is collapsed in this region. 
In a larger current system, localization lengths become even larger, 
and whole system is filled with extended states. 
QHE is broken down in this region. 
This is consistent with Kawaji et al.'s recent experiments and proposal.
\cite{f} 

In summary, we have shown that the anisotropic mean field states have 
unquantized Hall conductance, huge anisotropic longitudinal resistance, 
negative pressure, and negative compressibility. 
These electric properties are consistent with the recent experiments of 
the anisotropic compressible Hall state and of collapse phenomena. 
Negative pressure of these states does not lead to instability but instead 
leads to a formation of a narrow strip of compressible gas states if its 
density is low and formation of the bulk compressible gas with the 
depletion region if its density is around the half-filling. 
Consequently in the system of low density compressible gas states, 
the Hall resistance is kept in the exactly quantized value even though the 
longitudinal resistance is finite. 
This longstanding puzzle was solved from 
the unusual property of compressible Hall gas, namely negative pressure. 
Hence it plays important roles in the metrology of the QHE. 

Authors would like to thank Y. Hosotani and T. Ochiai for 
useful discussions. 
One of the present authors (K. I.) also thanks S. Kawaji and 
B. I. Shklovskii for fruitful discussions. 
This work was partially supported by the special 
Grant-in-Aid for Promotion of Education and Science in Hokkaido University 
provided by the Ministry of Education, Science, Sports, and Culture, the 
Grant-in-Aid for Scientific Research on Priority area (Physics of CP 
violation) (Grant No. 12014201), and the Grant-in-aid for 
International Science Research (Joint Research 10044043) from the Ministry 
of Education, Science, Sports and Culture, Japan.

\begin{figure}
\epsfxsize=2.5in\epsffile{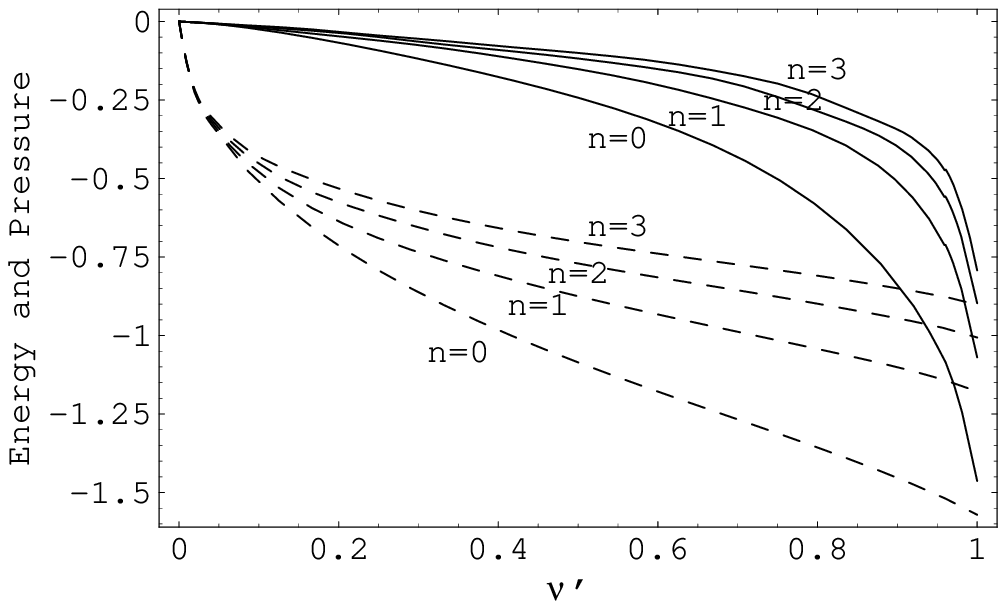}
Fig~1. Energy per particle (dashed lines) in the 
unit of $q^2/a$ and pressure (solid lines) in the unit of 
$q^2/a^3$ for $\nu=n+\nu'$, $n=$0, 1, 2, and 3. 
\end{figure}
\begin{figure}
\epsfxsize=2.4in\epsffile{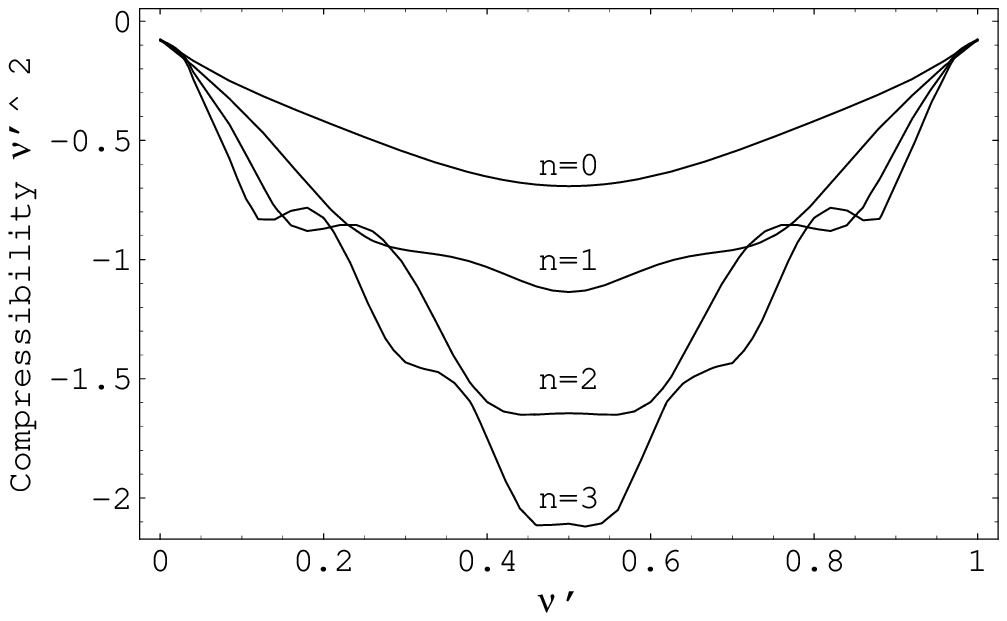}
Fig~2. Compressibility times $\nu'^2$ in the unit of $a^3/q^2$ 
for $\nu=n+\nu'$, $n=$0, 1, 2, and 3. 
\end{figure}
\begin{figure}
\epsfxsize=2.4in\epsffile{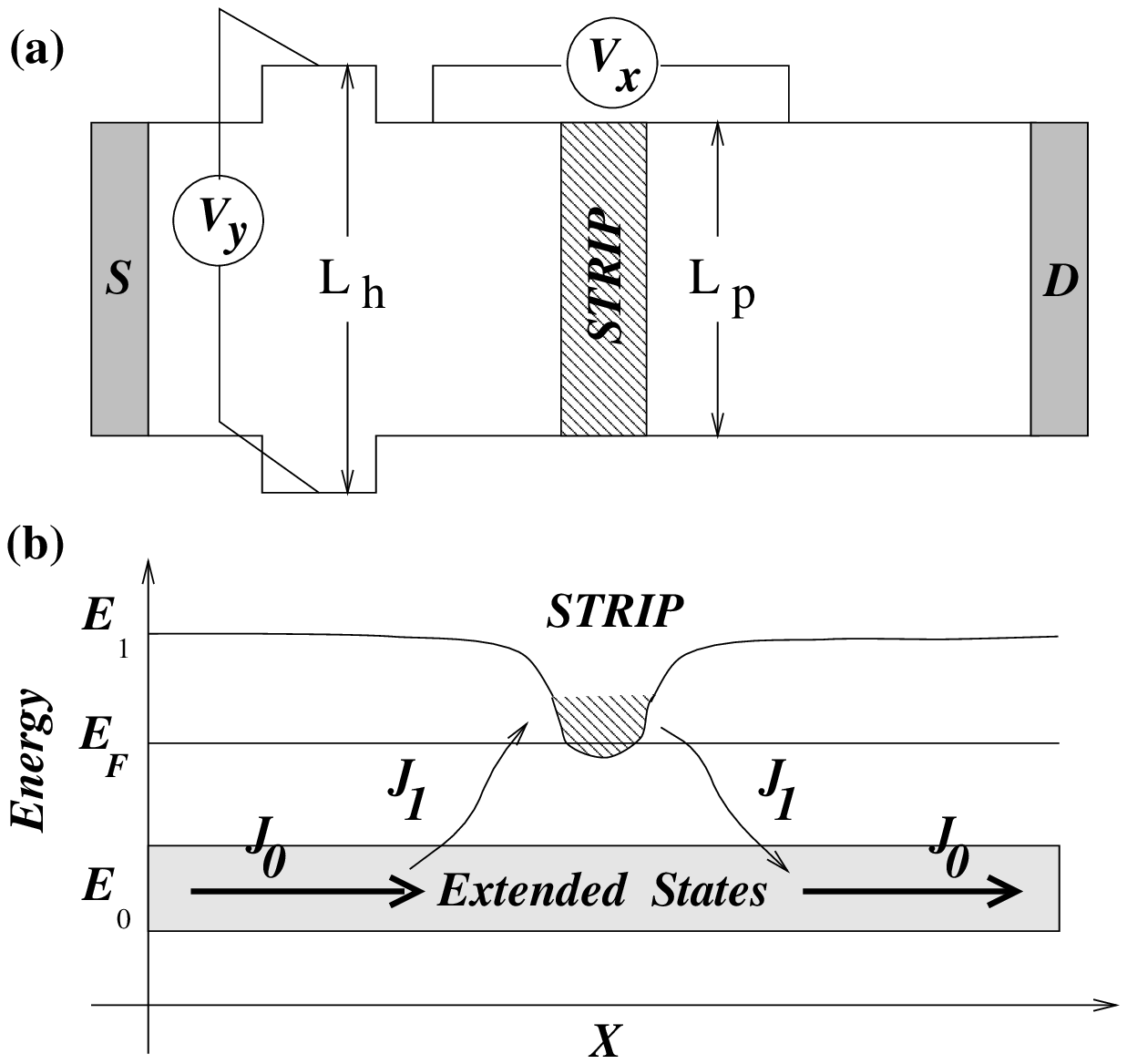}
Fig~3. (a) Schematical view of a Hall bar in the region II-(ii). 
Electric current is injected from S to D. 
(b) Sketch of the extended states carrying the undercurrent $J_0$ 
and the compressible state carrying the dragged current $J_1$ at 
the strip area in the energy and the x-position.  
\end{figure}
\begin{figure}
\epsfxsize=1.8in\epsffile{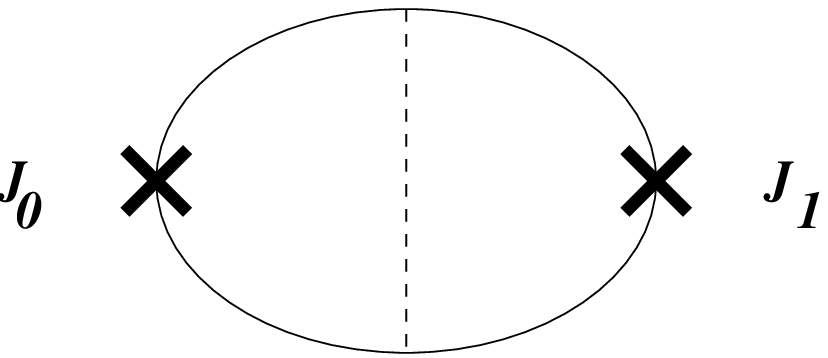}
Fig~4. Feynman diagram for the current-current correlation function. 
$J_0$ is the undercurrent carried by the extended states and 
$J_1$ is the dragged current carried by the compressible states in the 
strip area. The dashed line stands for an interaction effect. 
\end{figure}
\end{multicols}
\end{document}